\documentclass{jltp}

\usepackage{graphicx,amssymb,amsmath,amsbsy}
\usepackage[T1]{fontenc}
\def\slantfrac#1#2{\kern.1em^{#1}\kern-.3em/\kern-.1em_{#2}}

\def\m#1{\mathrm{#1}}
\newcommand{\greeksym}[1]{{\footnotesize{\usefont{U}{psy}{m}{n}#1}}}

\title{\bf{Search for supersolidity in \raisebox{0.8ex}{\bf{\footnotesize
        4}}He in low-frequency sound experiments.}}

\author{Yu.~Mukharsky, O.~Avenel, and E.~Varoquaux$^*$}
\address{ CEA-DRECAM, Service de Physique de l'\'Etat Condens\'e, \\
          Centre d'\'Etudes de Saclay, 91191 Gif-sur-Yvette Cedex (France)\\
      $^*$CNRS-Laboratoire de Physique des Solides, \\
          B\^at. 510, Universit\'e Paris-Sud, 91405 Orsay (France)}

\runninghead{ Yu. Mukharsky, O. Avenel, E. Varoquaux}{Search for supersolidity
in low-frequency sound experiments}

\begin{document}
\maketitle

\begin{abstract}
  We present results of the search for supersolid
  \raisebox{0.8ex}{\it{\footnotesize 4}}He using low-frequency, low-level
  mechanical excitation of a solid sample grown and cooled at fixed volume.
  We have observed low frequency non-linear resonances that constitute
  anomalous features. These features, which appear below $\sim$ 0.8 K, are
  absent in \raisebox{0.8ex}{\it{\footnotesize 3}}He.  The frequency, the
  amplitude at which the nonlinearity sets in, and the upper
  temperature limit of existence of these resonances depend markedly on the
  sample history.

\end{abstract}


\section{Introduction}

The observation of an anomalous behaviour in the moment of inertia of solid
$^4$He first reported by Kim and Chan\cite{Kim:04,Kim:04a} have recently been
confirmed by a number of different groups using the same torsional oscillator
technique, Keiya Shirahama {\it et al.} from Keio
University,\cite{Shirahama:06} Minoru Kubota {\it et al.} from the University
of Kyoto,\cite{Kubota:06} and Rittner and Reppy from Cornell
University.\cite{Rittner:06}. The last group however reports a marked
dependence of the supersolid response upon sample annealing.

Other experiments probing the mechanical properties of solid $^4$He at
temperatures below 1 K by a number of different techniques have concluded,
with increasing certainty as the techniques were refined, that no {\it dc}
superflow was taking place on a scale that would match that reported by Kim
and Chan (see \cite{Day:06} and references therein). A possible exception is a
recent gravitational flow experiment conducted on the liquid-solid coexistence
curve by Sasaki {\it et al.}\cite{Sasaki:06}

We report here measurements of the response of solid $^4$He samples to
low-frequency, low-level mechanical excitations that were carried out with the
goal of detecting the presence of an eventual supermobile component. It can be
expected that such a component would give rise to a hydrodynamic mode with a
lower propagation velocity than that of ordinary (first) sound. Such a mode,
if associated to a Bose condensate, should show little internal damping at low
temperature, and, in all likelihood, should display a critical velocity above
which the supermobile property breaks down.

\section{The experiment}

\begin{figure}[t]
  \begin{center}
    \includegraphics[width=7cm]{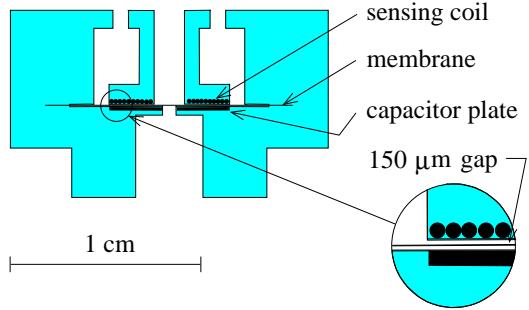}
    \caption{\label{cell} Schematic view of the experimental cell. The
      flexible membrane makes a partition between the bottom and the top
      chambers. As shown in the blown-up view, this membrane is separated by
      small gaps from the solid walls in which the flat electrodynamic pickup
      coil (top) and the capacitor counter-electrode (bottom) are embedded.}
  \end{center}
\end{figure}

The experimental cell used in these {\it ac} stress-strain measurements is
built from components of the hydromechanical resonator previously used for
phase slippage experiments in the $^4$He and $^3$He
superfluids.\cite{Avenel:87,Varoquaux:87} 

Pressure is applied to the solid by a flexible membrane positioned between two
solid walls as shown in Fig.\ref{cell}. The two flat cylindrical chambers
above and below the diaphragm have a height of approximately 150
{\greeksym{m}}m and a diameter of 8 mm. The top chamber is open to the main
volume of the cell by a 0.6 mm diameter cylindrical vent of height 0.5 mm. The
bottom chamber has a larger opening, 1.5 mm in diameter, 5 mm in length. The
path around the cell is $\sim$ 2.6 cm in length. The volume around the cell
has largest dimension 3.5 cm. The corresponding $\lambda/2$ resonance
frequency for a longitudinal sound velocity of 440 m/sec is 6.3 kHz, which
should fix the low end of the acoustic resonance spectrum in the cell.
 
Several flexible membranes have been used, made of 7.5 \greeksym{m}m thick
Kapton coated with a thin superconducting film of either Al or Nb. The
membrane is electrostatically actuated by applying between it and ground an
{\it ac} voltage of up to 7 volts rms superposed to a {\it dc} bias of
typically 150 volts. The corresponding force per unit area, of the order of 1
Pa, is much weaker than in other experiments probing solid
$^4$He.\cite{Day:06}

The displacement of the diaphragm is measured with an electrodynamic sensor
using a {\it dc}-SQUID as front-end amplifier to achieve a sensitivity of
$\sim 10^{-15}$~m, significantly higher than in other attempts to observe
supersolidity in $^4$He, apart possibly from torsional oscillator experiments.

\section{Experimental observations}

We have observed low frequency non-linear resonances that constitute anomalous
features in the sense that they appear below a sample-dependent onset
temperature $T_\m r$ and that their frequencies are too low to be usual
acoustic resonances. These resonances have not been observed in a pure solid
$^3$He sample. When the $^4$He samples, which are obtained by cooling along
the melting curve at constant volume, are grown and cooled very gently, these
resonances are either weak or even not seen.  Efforts to produce $^4$He
samples that would show reproducible features in a reproducible manner have so
far failed. In all likelihood, the solid sample ends up at low temperature
badly fractured, which greatly affects its low-frequency stress-strain
response.  Slowly raising and lowering the temperature while remaining in the
hcp phase changes, or even suppresses, these resonances.

\begin{figure}[t!]      \label{results}
  \begin{center} \mbox{
    \includegraphics[width=6.3cm]{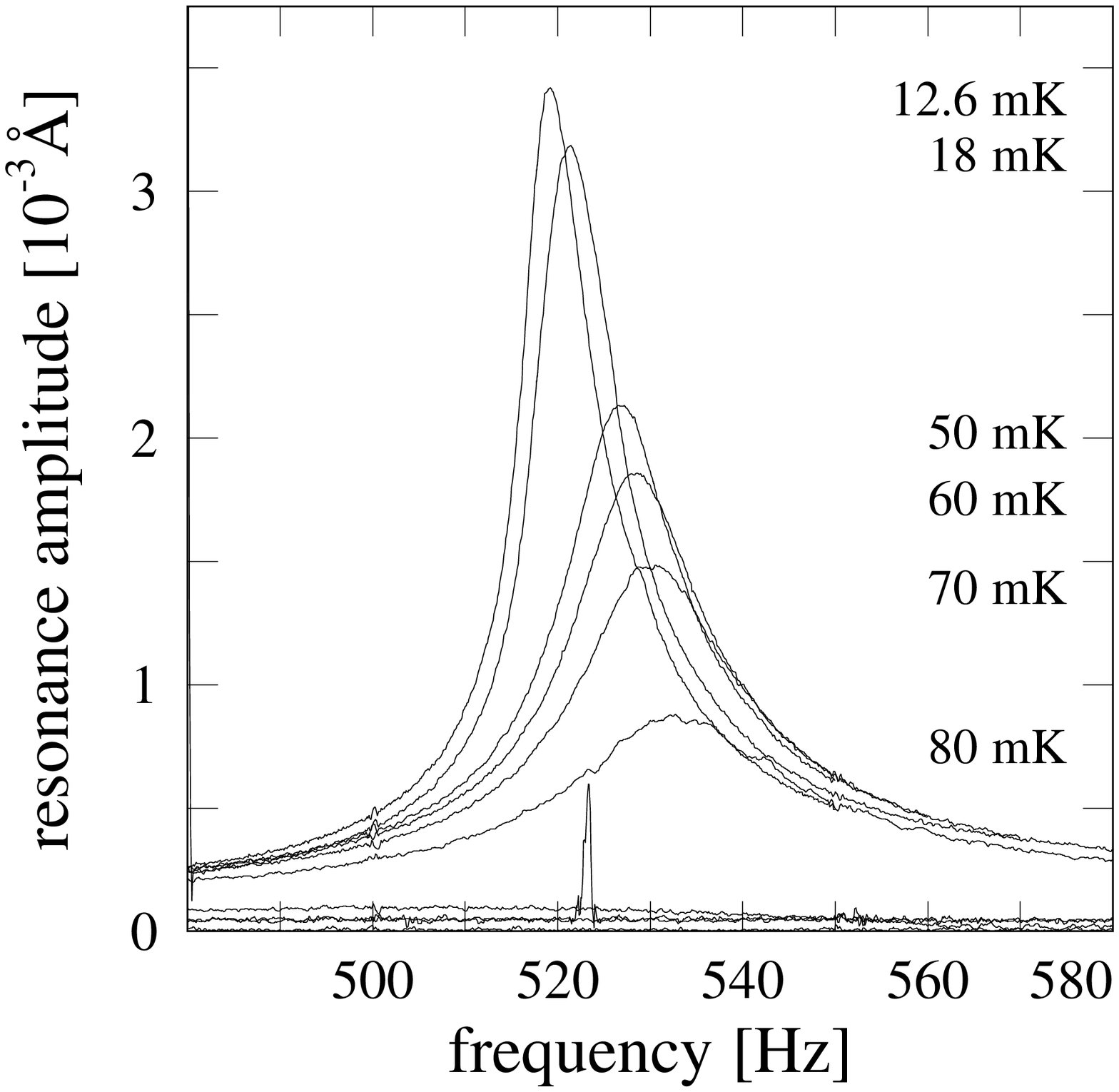}
    \includegraphics[width=6.3cm]{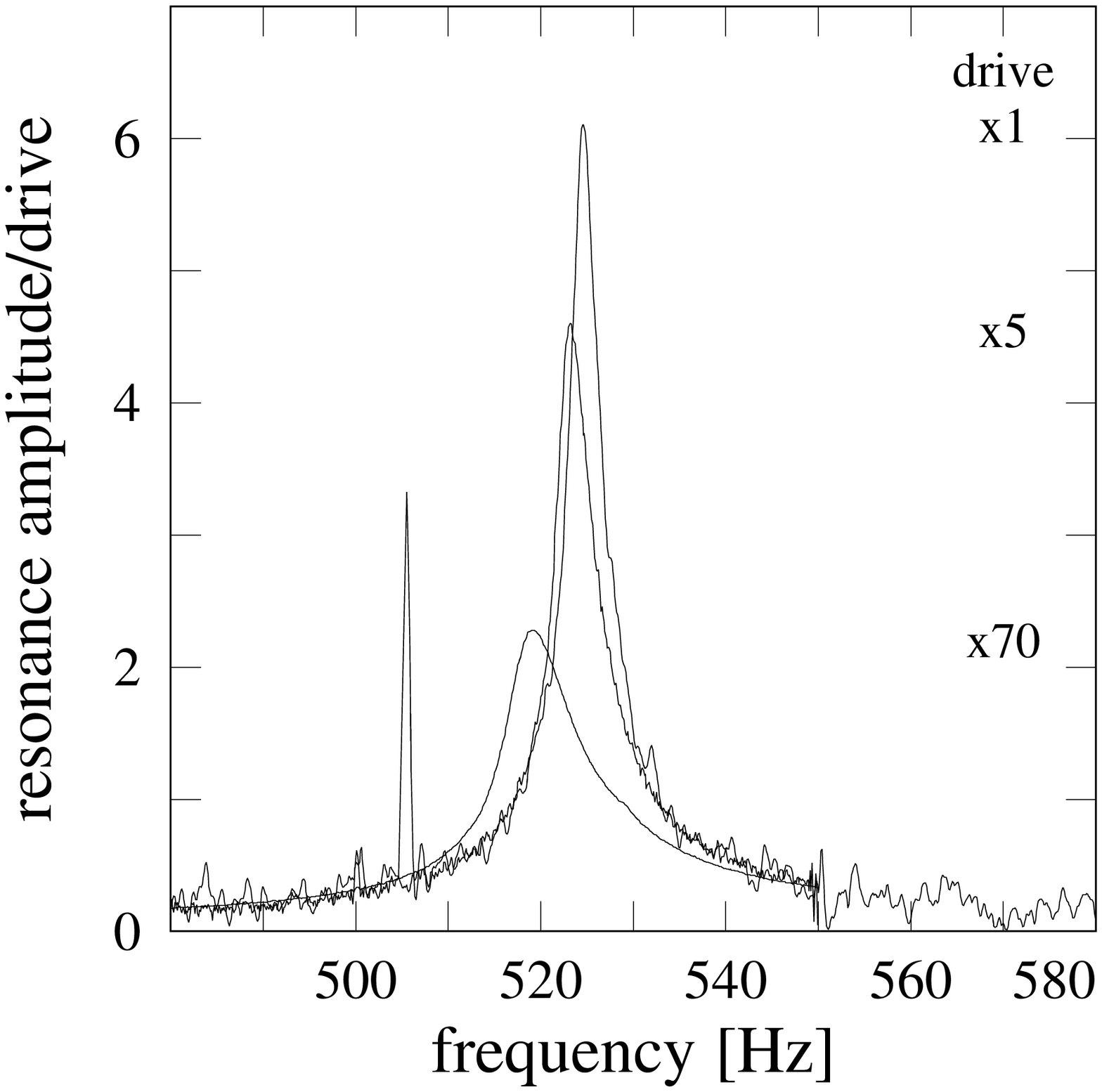}}
    \caption{(left) Resonance amplitude, corrected for the
      baseline, approximately in Angström with an uncertainty of less than
      30\%, {\it vs} frequency, in Hz, for several temperatures. The bottom
      curves are at 90, 100, 110, and 120 mK and no resonance can be separated
      from noise and baseline correction errors; (right) Resonance amplitude,
      corrected for the baseline, divided by drive amplitude in arbitrary
      units {\it vs} frequency, in Hz, for the relative drive levels
      $\times1$, $\times5$, and $\times70$. Sharp glitches due to spurious
      noise can be seen in both frames. The left frame is for the highest
      excitation level. }
  \end{center}
\end{figure}

Such a resonance is shown in Fig.\ref{results}
The displacement of the flexible membrane when the frequency is swept through
resonance at constant excitation drive level is plotted in terms of the
frequency (i) for various temperatures (left frame) at the $\times70$ drive
level; (ii) for various drive levels at 12.6 mK (right frame). When the
temperature is raised, the resonance amplitude decreases, the width increases
and the frequency shifts upward. At 90 mK, the resonance features are lost in
the baseline corrections uncertainties; the resonance disappears completely
above 90 mK.

When the drive level increases, the resonance amplitude also increases but not
in proportion to the drive, the resonance frequency shifts downward in
frequency, the width increases and the shape becomes distorted. The non-linear
behaviour does not appear to set in very abruptly. At low drive level and low
temperature, the resonance is fairly sharp, with a quality factor $Q$ of the
order of 100.

The data shown in Fig. \ref{results} were obtained after a fast cooldown from
1 K. The solid was formed by the blocked capillary method starting from a
pressure of 47 bars at a temperature of 2.3 K with nominal purity $^4$He
containing typically 1 part in $10^7$ of $^3$He impurities.  Besides the
resonance at 524 Hz shown in the figure, weaker resonances were also seen at
593, 808, 924, 1088, and 1392 Hz. These resonances are attached to a
particular sample configuration and usually remain unchanged as long as the
temperature is kept well below 1 K although some slow relaxation was
occasionally observed at low temperature. Raising the temperature above 1 K
and cooling again gives a different sample configuration, resulting in a
different set of resonance frequencies and a different onset temperature. Low
frequency resonances have been observed up to above 700 mK.  When cooling very
slowly from 1.5 K, these features are either weak or completely absent.

\section{Discussion}

The sample-history dependence of the low-frequency resonances is most probably
linked to the polycrystalline state that forms in the solid when it is cooled
in the fixed volume container. Thermal gradients develop and the solid sample
breaks under thermal stress. Annealing becomes very slow below $\sim$ 1.2 K
and the defects created during the cool-down remain frozen.

It has been known for some time \cite{Lie-zhao:86} that layers of superfluid
liquid persist at pressures up to 50 bars in very small pores of Vycor glass
while two amorphous solid layers are adsorbed on the glass surface and solid
forms in the bulk of the pores. This type of interfacial superfluidity has
been observed by Yamamoto {\it et al.}\cite{Yamamoto:04} in pores as small as
2.5 nm, in which superfluidity has been observed up to 35 bars at low
temperature.

It has been suggested by Beamish \cite{Beamish:04} and by Dash and Wettlaufer
\cite{Dash:05} that such superfluid layers may provide mechanical decoupling
between the $^4$He solid and the torsional oscillator walls and be a possible
explanation for experiments of Kim and Chan.\cite{Kim:04,Kim:04a} This slip
mechanism would however not explain why defects in the bulk of the crystal
sample appear to play an important role in the moment-of-inertia
anomaly.\cite{Rittner:06}

A number of authors have pointed out that a perfect crystal can not possess
off-diagonal long range order, a result established by Penrose and Onsager in
1952.\cite{Penrose:56a} It was then speculated that the superfluid layers
forming at the grain boundaries in a polycrystalline sample could play a major
role in the supersolidity.\cite{Burovski:05,Prokof'ev:05,Khairallah:05}

It seems quite possible that the stress-strain anomalies of solid $^4$He
formed at constant volume reported here are related to the existence of
pre-melted superfluid films at grain boundaries.  According to Dash and
Wettlaufer, the non-crystalline interface between grains may have a thickness
of 4$\sim$8 atomic layers.\cite{Dash:05} A fraction only of these layers forms
a 2D-superfluid, the rest being amorphous solid.  A Kosterlitz-Thouless
transition temperature $T_\m{KT}\sim0.1$ K in $^4$He film at saturated vapour
pressure would correspond to an equivalent superfluid fraction of 3\% of an
atomic layer.\cite{Rudnick:78} The critical velocity of such a dilute
2D-superfluid is in the range of a few cm/s.\cite{Gillis:89} If actual
superflow is thought to take place, this would imply that there are many
interfaces in parallel and that the grains are very small, in the range of 1
to 10 {\greeksym{m}}m in size. In such a case, the properties of this
solid-superfluid slurry would be more homogeneous than the wide variation of
resonance frequencies and onset temperatures would let think.

If there are indeed fewer crystallites, the pressure change induced by the
membrane can still be transmitted in the superfluid layers by fourth sound,
the velocity of which is, at low temperature, of the order of the first sound
velocity in the bulk.  In the very dilute Bose condensate, first sound
velocity is low.  It is expressed in terms of the scattering length $a_0$ and
the condensate number density $n_0$ by $c=(2\pi \hbar/m_4)\sqrt{n_0a_0/\pi}$,
$\hbar$ being the Planck constant and $m_4$ the atomic mass of
$^4$He.\cite{Mewes:96} Taking $a_0$ of the order of the hard core radius of
the $^4$He atomic potential (2.5 \AA) yields an estimate of 27 m/s for the
layer in which 3 \% of the atoms are in the condensate, a value low enough to
account for the observed resonances.  The frequency also depends on the
geometry of the actual path between grains.  So far, no systematic pattern of
behaviour for frequencies and critical temperatures has emerged in our
experiments.

To conclude, we have observed low frequency resonance modes that depend on the
presence of defects in the solid $^4$He sample. This observation brings
evidence that there exists a component of the sample that transmits low
velocity, low dissipation pressure waves through the sample. The superfluid
layers that are believed to coat the grain boundaries may constitute such a
component. 

We acknowledge useful correspondence with John Reppy and Sébastien Balibar.

\end{document}